# Thiolation and PEGylation of silicon carbide nanoparticles

Péter Rózsa[a,b], Olga Krafcsik[c,d], Zsolt Czigány[d], Sándor Lenk[c], David Beke[b,e], Adam Gali[b,c,f,*]

[a]*Hevesy György PhD School of Chemistry, Eötvös Loránd University, Pázmány Péter sétány 1/A, H-1117 Budapest, Hungary*

[b]*HUN-REN Wigner Research Centre for Physics, Institute for Solid State Physics and Optics, P.O. Box 49, H-1525 Budapest, Hungary*

[c]*Department of Atomic Physics, Institute of Physics, Budapest University of Technology and Economics, Műegyetem rakpart 3., H-1111, Budapest, Hungary*

[d]*HUN-REN Centre for Energy Research, Institute of Technical Physics and Materials Science, P.O. Box 49, H-1525, Budapest, Hungary*

[e]*Kandó Kálmán Faculty of Electrical Engineering, Óbuda University,Bécsi út 94-96, H-1034, Budapest, Hungary*

[f]*MTA-WFK Lendület "Momentum" Semiconductor Nanoparticles Research Group, P.O. Box 49, H-1525 Budapest, Hungary*

**gali.adam@wigner.hun-ren.hu*

**Abstract**

In this study, we implement thiol termination on the surface of few-nanometer-sized silicon carbide (SiC) nanoparticles (NPs) to enable further applications, such as fluorescent biomarkers. Various spectroscopic techniques are employed to monitor the effectiveness of the surface treatment. A thiol-Michael addition reaction is performed by conjugating 4-arm PEGmaleimide molecules to the thiol groups of SiC NPs, further demonstrating the reactivity of thiol-terminated SiC NPs, which also acts as a protection layer against oxidation. These fluorescent thiolated SiC NPs, both with and without conjugated molecules, are directly applicable as bioinert probes. Since SiC NPs can potentially host room-temperature fluorescent defect quantum bits, our results are an important step to realize a bioinert, ultrasmall quantum sensor bioagents, which may open new avenues in biotechnology.

## 1. Introduction

Nanoparticles offer immense potential across a broad spectrum of applications, including energy, environment, information technologies, agriculture, and biomedicine [1]**.** Semiconductor nanoparticles, in particular, are notable for their photocatalytic properties [2,3], and tunable fluorescence [4–6]. When incorporated into polymer matrices, the resulting composites demonstrate enhanced mechanical strength, electrical conductivity, and radiation absorbance. [7–13] They are also known for their utility in photovoltaics and sensing. These advantages are rooted in quantum confinement effects and high surface-to-volume ratios. Materials composed of earth-abundant and non-toxic elements further increase the appeal of these nanoparticles, making them suitable for biomedical use.

* Corresponding author. Tel.: +36-1-392-2222; e-mail: gali.adam@wigner.hun-ren.hu.

Among these, silicon carbide (SiC) has emerged as a promising multifunctional material. In addition to its mechanical hardness and chemical stability [14,15], SiC is a host for optically addressable point defect centers [16,17], including spin qubits [18–20], positioning it at the frontier of solid-state quantum technologies [17,21–23]. Moreover, SiC nanoparticles (NPs) exhibit low toxicity [24,25], are biocompatible, and—when reduced to <10 nm—show quantum emission suitable for imaging applications [26–28]. Incorporating SiC NPs into various matrices enhances material performance, including improving the mechanical strength and electrical conductivity of polymer nanocomposites [10,29–34]. These properties make SiC NPs ideal candidates for combining quantum sensing and biomedical functionalities in a single platform.

A critical factor limiting broader adoption is the need for effective surface functionalization, as the native surface chemistry of SiC NPs—rich in oxygen-containing moieties—determines their colloidal stability, biological interactions, and optical properties [35–39]. Engineering the surface with functional groups such as amines, carboxyls, and thiols enables chemical conjugation, bio-recognition, and sensing capabilities. While prior work has focused on amine-functionalized SiC NPs [40,41], thiol-terminated SiC NPs have yet to be reported, despite the broad applicability of thiols in biosensing, click chemistry, and nanoparticle stabilization [42–48].

Here, we introduce thiol-functionalized SiC NPs synthesized via aqueous-phase chemistry, inspired by methods previously used for nanodiamonds [49,50]. To overcome the inherent challenges of modifying ultrasmall (~3 nm) SiC NPs—especially in the presence of surface silicon sites—we optimized a thiolation protocol and confirmed successful functionalization using FTIR, XPS, and thiol-Michael addition via maleimide conjugation. Importantly, we demonstrate that the resulting SiC-SH NPs preserve favorable optical properties, which are crucial for applications in imaging and quantum sensing.
Given the high sensitivity of thiol groups to oxidative degradation, we further stabilized the functionalized surface by PEGylation using maleimide-terminated polyethylene glycol (PEG). PEGylation not only protects the reactive thiol groups from oxidation—as supported by previous studies [51–54]—but also enhances the biocompatibility, colloidal stability, and functional integration of the nanoparticles. This dual modification strategy provides a robust route for biofunctionalization, enabling conjugation to biomolecules and use in targeted biomedical applications. Furthermore, PEGylation [42,51,55–62] extends the potential of SiC-SH NPs for use in drug delivery, tissue engineering, and as platforms for quantum biosensing [63,64].

This study reports stable thiol functionalization of SiC NPs and demonstrates the synergistic benefits of PEGylation for preserving functionality and enhancing biointegration. Together, these findings establish a foundation for utilizing surface-engineered SiC NPs that incorporate defect qubits in future biomedical and quantum technology applications.

## 2. Materials and methods

### *2.1. Materials*

The following materials were employed in this study: hydrofluoric acid (HF) (GPR Rectapur, VWR Chemicals), nitric acid ($HNO_3$) (GPR Rectapur, VWR Chemicals), ammonia solution (28%) (GPR Rectapur, VWR Chemicals), sodium borohydride ($NaBH_4$) (>98% powder, Sigma-Aldrich), hydrobromic acid (HBr) (48%, Acros Organics), acetic acid (HOAc) (96%, Reanal), thiourea (Reanal), sodium hydroxide (NaOH) (Analar Normapur, pellets, Sigma-Aldrich), sulfuric acid ($H_2SO_4$) (95%, Normapur, VWR Chemicals), and 4-arm-PEG-maleimide-10k (Sigma-Aldrich). All chemicals were used as received without further purification.

### *2.2. Instruments and characterizations*

Conductivity measurements were done by using a CO 30 VWR conductivity meter.

FTIR spectra were acquired using a Bruker IFS 66v/S spectrometer operating at 3 mbar pressure, equipped with a Globar source and liquid nitrogen-cooled MCT (mercury cadmium telluride) detector. The samples were drop-cast on a single-side polished <100> Si wafer.

Photoluminescence (PL) spectra were recorded using Horiba Jobin-Yvon Fluorolog FL3-22 spectrophotometer equipped with a 450 W Xenon lamp, an iHR-320 grating monochromator and an R928 photomultiplier tube. The samples were measured in a 10 × 10 × 40 mm quartz cuvette (Hellma 110QS, Jena, Germany).

Atomic force microscopy (AFM) imaging was performed using a Bruker Dimension Icon with Bruker MPP-111, 00-10 (Leipzig, Germany) probe equipment in tapping mode. 10 μL samples were drop-cast in a single-side polished <100> Si wafer and allowed to evaporate at room temperature. The scanning was processed around the dried droplets. 25 $\mu m^2$ area was scanned, and the height of at least 100 particles was measured for size distribution.

XPS measurements were conducted using a twin-anode X-ray source (Thermo Fisher Scientific, Waltham, MA, USA, XR4) and a hemispherical energy analyzer with a nine-channel multi-channeltron detector (SPECS-GROUP, Berlin, Germany, Phoibos 150 MCD). The base pressure in the analysis chamber was approximately 2 × $10^{-9}$ mbar. Mg $K_\alpha$ radiation (1253.6 eV) was used without monochromatization. Samples were drop-cast onto a niobium (Nb) substrate (GoodFellow GmbH, Hamburg, Germany). All spectra were corrected using the Nb $3d_{5/2}$ orbital of $Nb_2O_5$ at 207.1 eV and the C1s peak of adventitious carbon (C–C/C–H) at 284.8 eV. For the S2p orbital, a 1:2 ratio of $S2p_{1/2}$:$S2p_{3/2}$ and an energy separation of 1.2 eV were used as split parameters. Peak fitting was performed using a GL30 function (Gaussian/Lorentzian, 30% Lorentzian character) with Shirley-type background subtraction.

High-resolution transmission electron microscopy (HRTEM), energy-dispersive spectroscopy (EDS), and electron diffraction analyses were performed using a Thermis TEM (Thermo Fisher) operated at 200 kV and equipped with spherical aberration (Cs) correction in the imaging system, providing a spatial resolution of 0.8 Å in HRTEM mode. Images were acquired using a 4 × 4 k Ceta camera with Velox software (Thermo Fisher). Samples were drop-cast onto a copper TEM grid (TED Pella) coated with an ultrathin lacey carbon layer and dried under an infra-LED lamp.

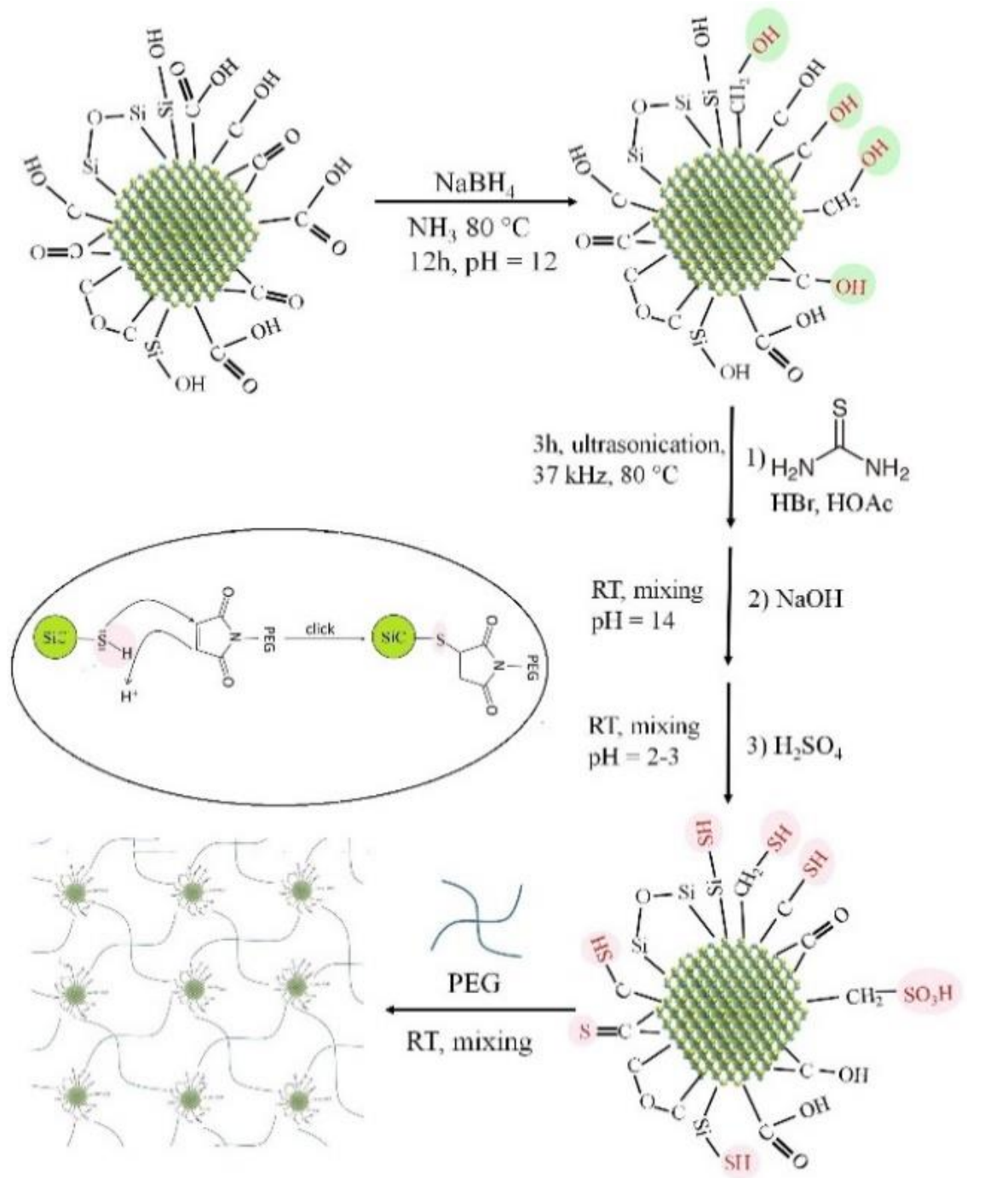

Figure 1. Synthesis of thiolated and PEGylated SiC NPs.

*2.3. Synthesis of thiolated and PEGylated SiC NPs*

SiC NPs were synthesized using the NPEGEC etching method [18,35,39,40,65–68]. Briefly, 5 g of SiC microcrystals were etched in 40 mL of a 3:1 mixture of concentrated HF and $HNO_3$ at 150 °C in a PTFE-lined hydrothermal reactor. This process generated mesoporous layers on the microcrystals, which were subsequently disaggregated into nanocrystals via ultrasonication in an aqueous solution. The resulting nanocrystals were separated from the remaining microcrystals by centrifugation, yielding a colloidal solution of SiC NPs. The as-prepared SiC NPs are referred to as SiC-COOH.

The thiol-terminated SiC NPs were prepared in two steps. First, mild surface reduction was performed on SiC-COOH NPs (2 mg/mL) The pH of the NP suspension was adjusted to 14 by adding a few drops of $NH_3$ (25%) solution. Subsequently, $NaBH_4$ (0.08 g for 10 mL of NP suspension) was added, and the mixture was stirred at 80 °C for 12 hours. We refer to these reduced samples as SiC-OH. To create thiol-terminated NPs, 10 mL of a 2 mg/mL aqueous SiC-OH solution was placed in a 25 mL round-bottom flask and evaporated to dryness. Next, 3 mL of an HBr–HOAc solution (1:2 volume ratio) and 0.11 g of thiourea were added. The mixture was ultrasonicated at 80 °C for 3 hours at 37 kHz (Elmasonic P ultrasonicator). Then, 6 mL of a cold NaOH solution (7.5 M) was added to the hot mixture in an ice bath, followed by stirring at room temperature for 12 hours. To reduce the pH to 1–2, 96% $H_2SO_4$ was gradually added to the solution. Because of the high salt content precipitating during this step, the suspension was subsequently diluted and purified using a 1 kDa Pall Macrosep centrifugal filter until the permeate conductivity reached 0 µS. We label the thiol-functionalized samples as SiC-SH.

PEGylation was carried out by adding 3 mg of 4-arm maleimide–PEG to 0.5 mL of the SiC-SH suspension, followed by 12 hours of stirring. The mixture was then purified using a 1 kDa Pall Macrosep centrifugal filter. The resulting samples are referred to as SiC-S-PEG (Figure 1).

## 3. Results and discussion

The size and shape of the NPs before and after surface modification were determined by HRTEM (Figure 2) and AFM (Figure 3). The mean diameters derived from the AFM height profiles were 4.1 nm, 2.8 nm, and 5.4 nm for SiC-COOH, SiC-SH, and SiC-PEG, respectively. We attribute the observed size reduction following surface modification to the purification process, in which a substantial portion of the larger particles may have been removed. Electron diffraction performed on an assembly of SiC-SH particles confirmed that the material retained its SiC composition.

Notably, SiC-PEG exhibited a broader size distribution, and AFM images in Figure 3(c) revealed the presence of flat, wide structures. These results are consistent with HRTEM analyses presented in Figure 2(b). After PEGylation, the NPs tended to form larger structures (~30 nm), although the measured heights in AFM still corresponded to the original SiC-SH NPs. These findings suggest that the NPs are embedded in a soft polymer matrix, forming flattened, extended structures.

We also measured the polymer alone [Figure 3(d)], which has a mean size of ~1.6 nm, consistent with literature values for 10 kDa PEG molecules [69] that form a roughly spherical 3D structure.

The FTIR spectrum (Figure 4) of SiC-COOH NPs,

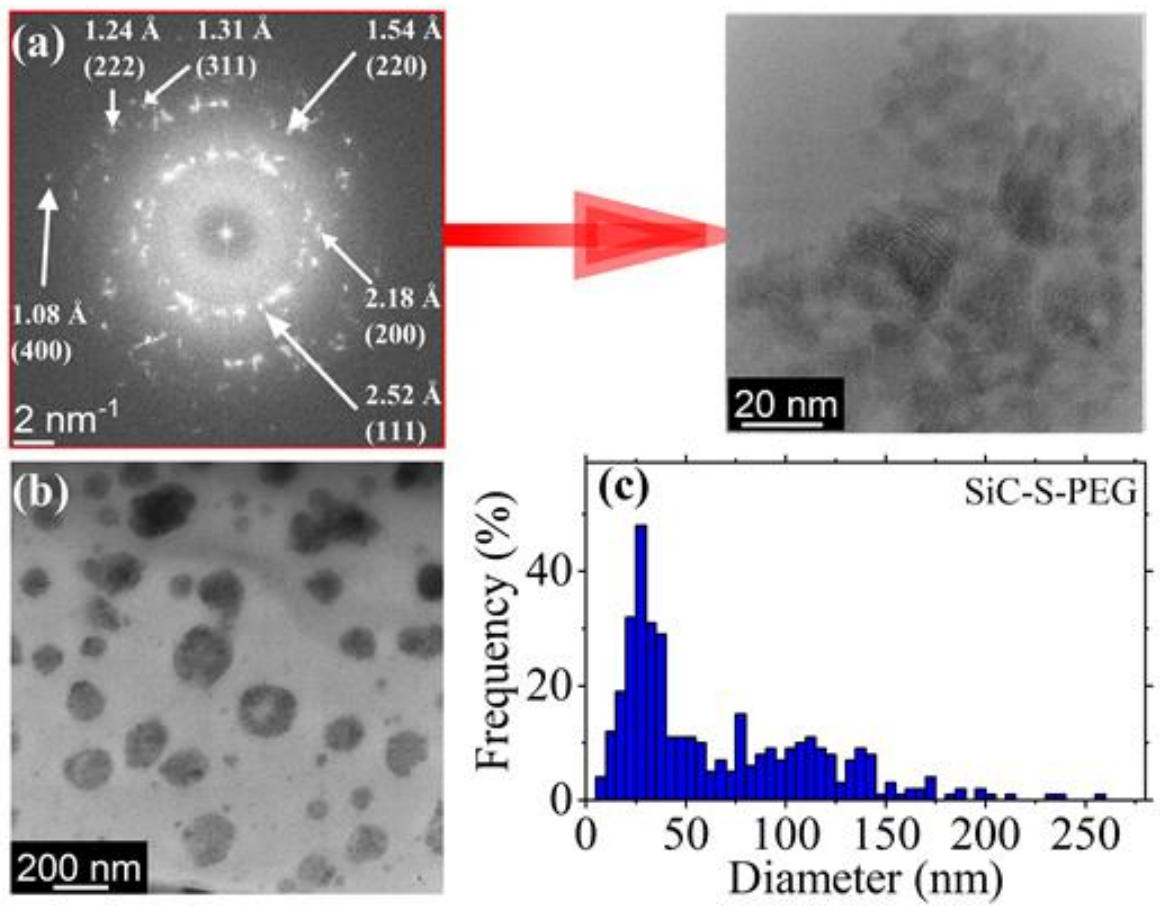

Figure 2. FFT image of SiC-SH NPs (a) measured on an ensemble of SiC-SH NPs (b), HRTEM image of SiC-S-PEG (c), the size distribution of SiC-S-PEG (d). Size distribution was calculated by measuring the diameters of 250 NPs from several areas.

SiC-SH NPs and PEG of SiC-COOH NPs primarily shows oxygen-related vibrational modes, including Si–O–Si/Si–O–C/C–O–C near 1100 $cm^{-1}$, –OH stretching around 3000 $cm^{-1}$, and C=O (carboxyl and other carbonyl) between 1600 and 1750 $cm^{-1}$ [70]. After thiolation, significant changes were observed: the relative intensities of the C=O, COOH, and OH bands decreased markedly. In contrast, C–S, –S–H, and –S–S vibrations typically exhibit weak FTIR signals and were not detectable. A peak initially at 1437 $cm^{-1}$ shifted to 1420 $cm^{-1}$ (indicated with a dashed line), it may be associated with various overlapping vibrations (e.g., C–H, OH, sulfate). The notable reduction in OH content likely modifies hydrogen bonding, resulting in peak shifts and narrower features. However, partial oxidation of thiol groups to organic sulfates cannot be ruled out.

The FTIR spectrum of SiC-S-PEG shows characteristic peaks attributed to maleimide–PEG molecules [71–74], effectively masking the vibrations

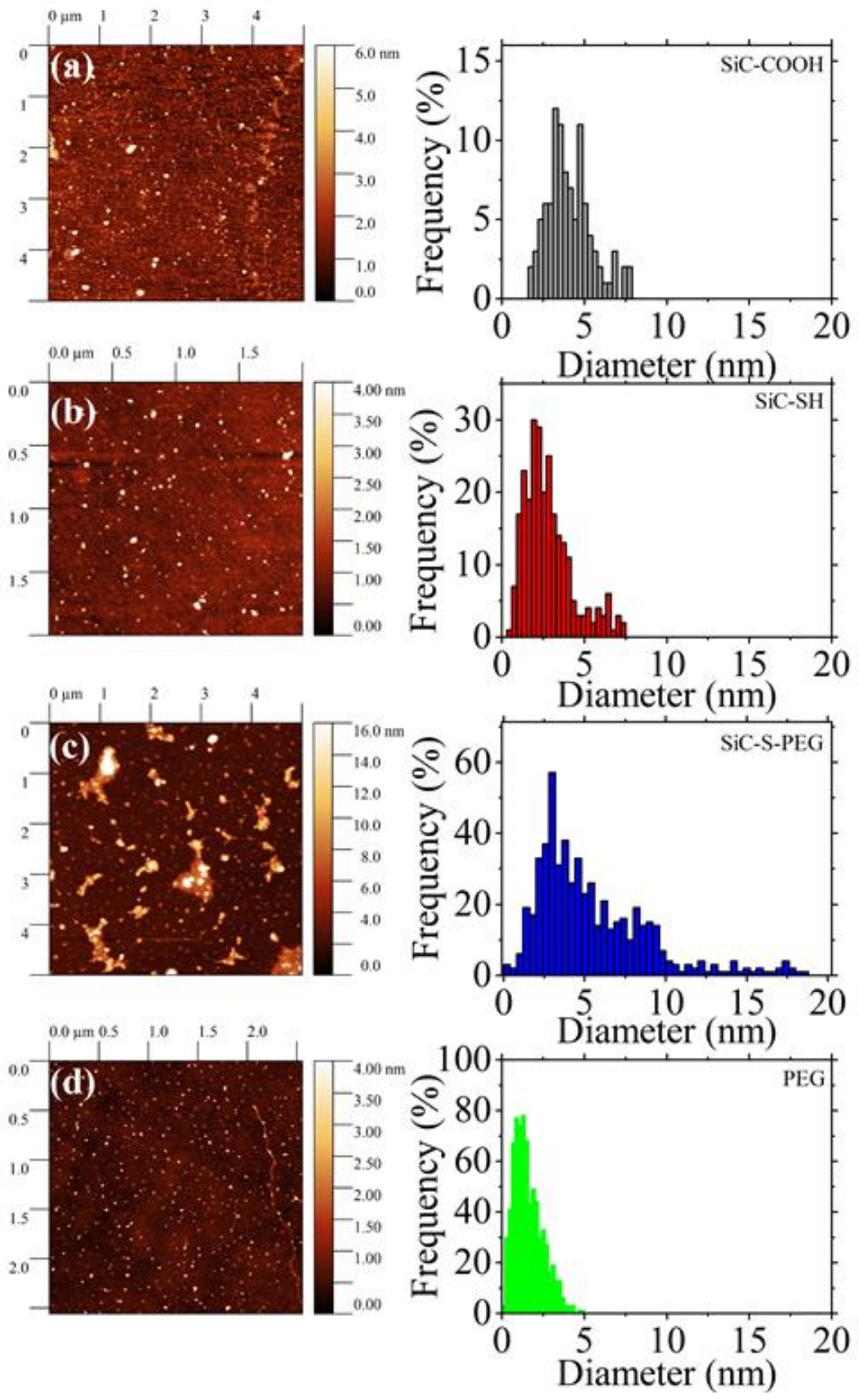

Figure 3. AFM images and size distributions of SiC-COOH (a), SiC-SH (b), SiC-S-PEG (c), and PEG (d) NPs.

of other surface groups and rendering them undetectable by FTIR analysis [73].

Survey XPS spectra and EDS analysis confirmed an increase in sulfur content following thiolation (Table 1).

Table 1. Elemental analysis of SiC-COOH, SiC-SH and SiC-S-PEG samples from XPS analysis. The values are in atomic concentrations (%).

| sample | C | Si | O | S | N |
|---|---|---|---|---|---|
| SiC-COOH | 51.3 | 3.5 | 44.8 | 0.1 | 0.3 |
| SiC-SH | 46.8 | 4.3 | 43.5 | 4.4 | 1.1 |
| SiC-S-PEG | 64.7 | 4.0 | 30.3 | 0.1 | 0.9 |

High-resolution C1s spectra also indicated a complex surface chemistry, with contributions from C–Si, C–C, C–O, C=O, and C–F in SiC-COOH NPs as plotted in Figure 5(a). Notably, the binding energies for C=O/C=S and C–O/C–S can overlap [75,76]. Following thiolation, the intensities of the O–C=O and C=O/C=S peaks decreased, while the C–O/C–S signal increased—consistent with the formation of C–S or C–O–S moieties, as supported by FTIR data showing a reduction in OH content.

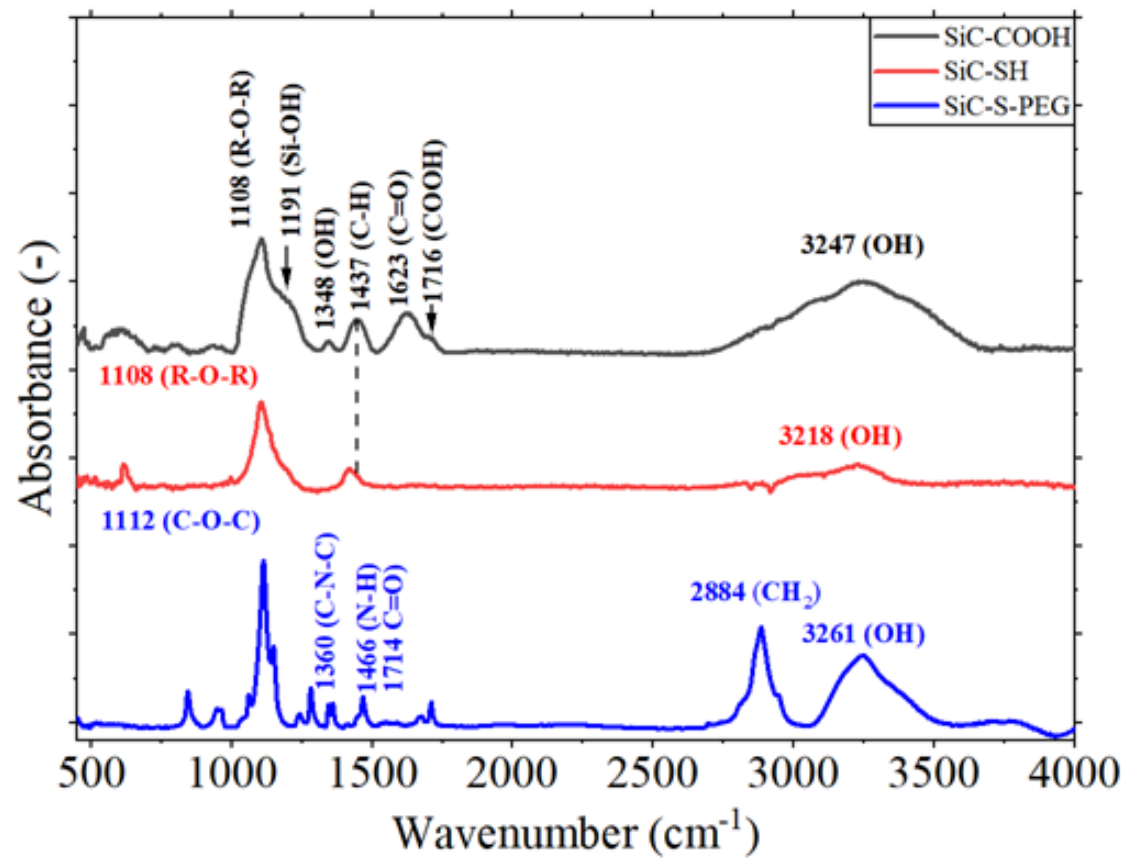

Figure 2. FTIR spectra of SiC-COOH (a), SiC-SH (b), and SiC-S-PEG (c) NPs. Spectra are vertically shifted for clarity

In the C1s region of SiC-S-PEG, the peak structure aligns with PEG spectra reported in the literature, displaying C–O/C–N, C=O, and O=C–N–H components [66]. However, the relative intensity of C–O compared to C–C increased, consistent with findings where PEG is used as a protective agent [51].

.

Table 2. Peak fitting parameters and component intensities of C1s photoelectron spectra of SiC-SH and SiC-S-PEG NPs. B.E. means binding energy, I.r. means intensity ratio.

| sample | O-C=O | | C-O/C-N C=O/C=S/ O=C-N-H | | C=O/C=S | | C-O/C-S | | C-C/C-H | | Si-C | |
|---|---|---|---|---|---|---|---|---|---|---|---|---|
| | B.E. (eV) | I.r. (%) | B.E. (eV) | I.r. (%) | B.E. (eV) | I.r. (%) | B.E. (eV) | I.r. (%) | B.E. (eV) | I.r. (%) | B.E. (eV) | I.r. (%) |
| SiC-COOH | 290.1 | 11.6 | - | - | 288.3 | 21.3 | 286.3 | 9.1 | 284.8 | 50.4 | 283 | 1.5 |
| SiC-SH | 289.6 | 6.2 | - | - | 288.1 | 11 | 286.4 | 15.2 | 284.8 | 53.9 | 282.7 | 2.1 |
| SiC-S-PEG | - | - | 287.6 | 6.6 | 288.6 | 2.2 | 286.4 | 70.4 | 284.8 | 20.9 | | |

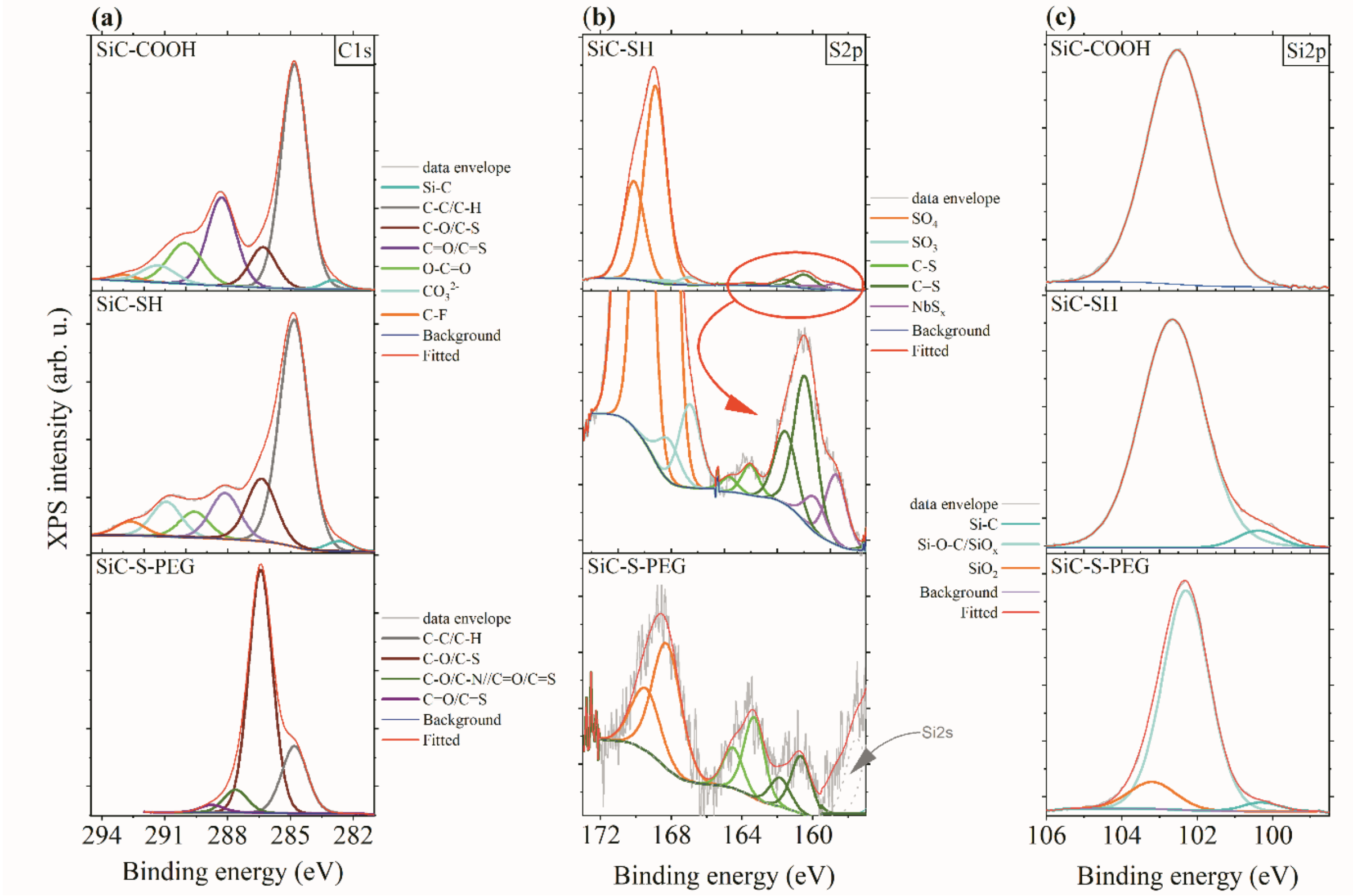

Figure 5. (a) C1s, (b) Si2p, (c) S2p photoelectron spectra of SiC-COOH, SiC-SH, SiC-S-PEG NPs, respectively

The S 2p spectra for SiC-COOH, SiC-SH, and SiC-S-PEG indicate that a significant fraction of sulfur in SiC-SH was present in oxidized form (e.g., sulfates, O–S–O), although C–SH and C=S signals are also detectable as shown in Figure 5(c). Thiol groups can be readily oxidized, which explains the relatively low concentration of free thiols. The presence of C=S may result from thione–thiol tautomerism [77,78].

Table 3. Peak fitting parameters and component intensities of S2p$_{3/2}$ photoelectron spectra of SiC-SH and SiC-S-PEG NPs. B.E. means binding energy, I.r. means intensity ratio

| sample | $SO_4$ | | $SO_3$ | | SH | | C=S | | $NbS_x$ | |
|---|---|---|---|---|---|---|---|---|---|---|
| | B.E. (eV) | I.r. (%) | B.E. (eV) | I.r. (%) | B.E. (eV) | I.r. (%) | B.E. (eV) | I.r. (%) | B.E. (eV) | I.r. (%) |
| SiCSH | 168.9 | 87 | 167 | 3.4 | 163.5 | 0.9 | 160.4 | 6 | 158.7 | 2.7 |
| SiC-S-PEG | 168.3 | 56.2 | - | - | 163.3 | 25.4 | 160.7 | 18.4 | - | |

In SiC-S-PEG, oxidized sulfur species decrease, and the tautomeric equilibrium shifts toward thiol. Maleimides form stable thioether linkages with free thiols, and thioether binding energies can be comparable to those of thiols. Additionally, the PEG environment likely provides protection against oxidation, explaining both the observed shift in peak ratio and the reduction in oxidized sulfur.

Interestingly, an additional low-energy peak (<160 eV) was observed in the S2p region of SiC-SH but disappeared after PEGylation. While the S2p binding energies are usually higher than 160 eV, peaks at unusually low binding energies have been documented for S–Nb bonds [79,80]. We speculate that the niobium substrate used in the XPS analysis might form bonds with unprotected thiols, explaining the appearance of this low-energy peak and its disappearance upon PEGylation [81,82].

The Si2p spectra revealed three distinct bonding environments [Figure 5(b)]. In SiC-COOH NPs, only silicon–oxicarbide signals were consistently observed, in agreement with previous studies [40,83]. Upon thiolation (SiC-SH), the appearance of the Si–C peak suggested partial removal of the surface oxide layer. After PEGylation (SiC-S-PEG), a $SiO_2$ peak emerged, likely due to surface oxidation during the Michael addition reaction. Although Si–S bonds typically appear around ~102 eV, fitting the Si2p spectra using Si–C, Si–O, and Si–Ox components yielded a good match, with no significant increase in oxidized Si species after thiolation. This suggests that Si sites were not extensively involved in the thiolation reaction.

Table 4. Peak fitting parameters and component intensities of Si2p photoelectron spectra of SiC-SH and SiC-S-PEG NPs. B.E. means binding energy, I.r. means intensity ratio

| Sample | $SiO_2$ | | Si-O-C/$SiO_x$ | | Si-C | |
|---|---|---|---|---|---|---|
| | B.E. (eV) | I.r. (%) | B.E. (eV) | I.r. (%) | B.E. (eV) | I.r. (%) |
| SiC-COOOH | - | - | 102.5 | 100 | - | - |
| SiC-SH | - | - | 102.6 | 94.4 | 100.3 | 5.6 |
| SiC-S-PEG | 103.3 | 22.1 | 102.2 | 75.3 | 100.3 | 2.6 |

Combining elemental analysis, FTIR, and XRD confirms successful surface modification of SiC NPs. The thiolated SiC surface reacts with maleimide groups, leading to PEGylated NPs that form a soft, polymer-based composite and protect residual thiol groups from further oxidation. We examined the photoluminescence (PL) of the NPs to assess the impact of this new surface chemistry on the optical behavior of luminescent SiC NPs (Figure 6). It is known that quantum confinement effects in SiC NPs are dominant for particle sizes above ~3 nm [68], while smaller NPs exhibit surface-related emissions. This emission partly originates from the HOMO (Highest Occupied Molecular Orbital) of the surface groups, which vary with surface termination [39]. Consequently, thiolation in the size regime studied here is expected to alter the PL spectrum compared to as-prepared SiC-COOH NPs.

Indeed, the PL of SiC-SH NPs was blue-shifted, similar to SiC-OH NPs (peak maximum at 435 nm) [39] although the shift was somewhat smaller. We attribute this difference to the higher electronegativity of –OH groups relative to –SH, wherein strongly negatively polarized –OH groups shift the HOMO downward more substantially. During PEG conjugation, SiC-S-PEG exhibited a further blueshift and a broader emission spectrum relative to SiC-SH, attributable to the formation of thioether bonds. These thioethers feature more negatively polarized sulfur than –SH, again lowering the HOMO, as well as possible contributions from PEG itself. The pronounced blue shift after PEGylation highlights the sensing potential of SiC-SH NPs, since sulfur bridge formation could be exploited when the –SH surface is used for target sensing.

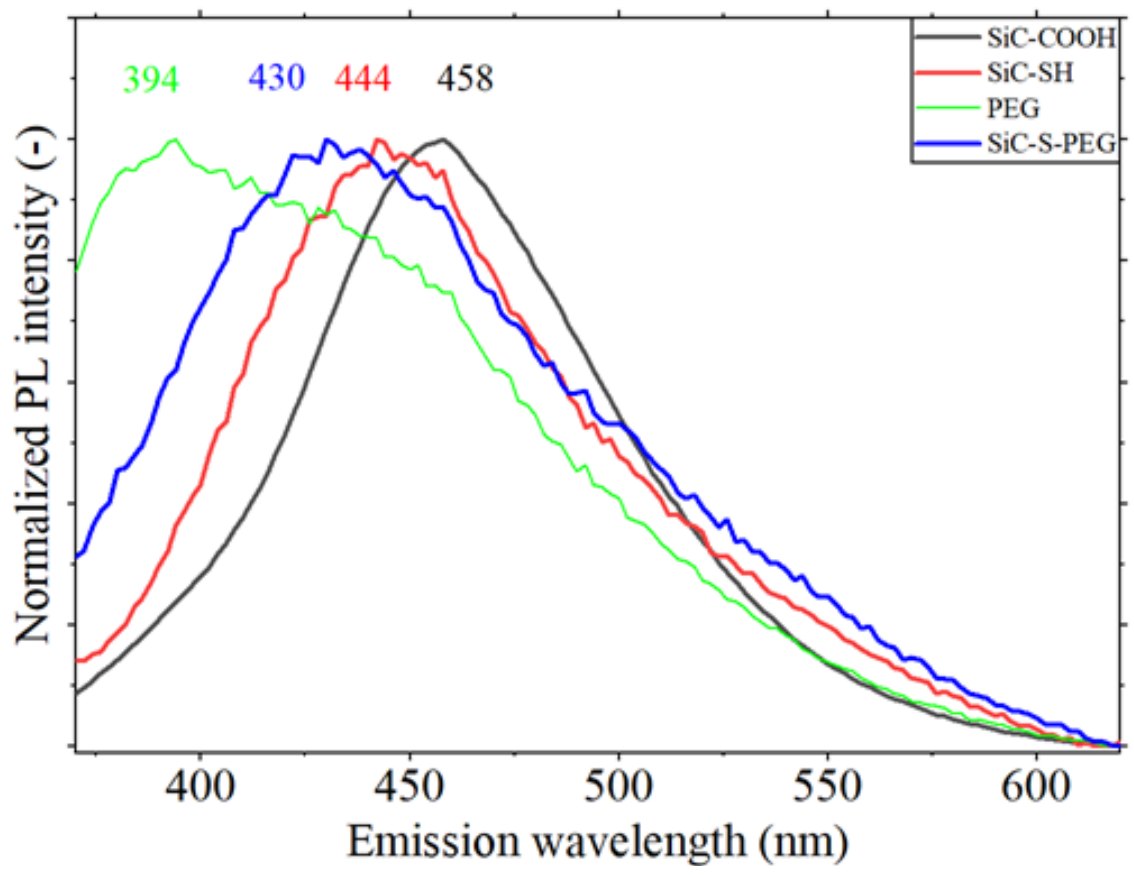

Figure 6. PL spectra of SiC-COOH (a), SiC-SH (b), PEG (c) and SiC-S-PEG (d) NPs

## 4. Conclusions

In this work, we synthesized SiC NPs and demonstrated their thiolation and PEGylation as an effective strategy for tunable surface engineering. From the XPS results, we have concluded that the PEG molecule serves not just as a reactant but also as a protective agent. These modifications significantly affected the photoluminescence properties of SiC NPs, indicating that both –SH groups and subsequent PEG conjugation can modulate optical emission by altering the nanoparticles' surface chemistry. Our study provides a pathway for integrating defect qubits in surface-modified SiC NPs for use in biological and quantum technology applications. The thiol termination is crucial, providing a versatile reactive platform for bioconjugation, which yields hydrophilic, colloidally stable, and biocompatible nanoprobes. Such nanocrystals are promising for targeted bioimaging and nanoscale quantum sensing [63], through, for example, the disulfide-thiol redox sensitive mechanism [64] applications, enabling long-term cellular imaging and quantum magnetometry *in vivo*.

## Data Availability Statement

All data available from the authors open reasonable requests.

## Author Contributions

Conceptualization, D.B. and A.G.; methodology, P.R., O.K., S.L., Z.C. and D.B.; validation, D.B. and A.G.; formal analysis, O.K., P.R. and D.B.; data curation, P.R. and O.K.; funding acquisition, A.G.; investigation, P.R., O.K., and S.L.; project administration, A.G.; resources, A.G.; supervision, A.G.; writing—original draft, P.R. and A.G.; writing—review and editing, D.B., P.R. and A.G. All authors have read and agreed to the published version of the manuscript.

## Funding

This study was supported by the Quantum Information National Laboratory sponsored by National Research, Development and Innovation Fund of Hungary (NKFIH) Grant No. 2022-2.1.1-NL-2022-00004 for the XPS and AFM investigations. The materials preparation was supported by the Ministry of Culture and Innovation of Hungary from NKFIH, financed under the TKP2021-NVA funding scheme (Project no. TKP2021-NVA-04). The HRTEM investigation was supported by the grant no. VEKOP-2.3.3-15-2016-00002 of the European Structural and Investment Funds. S. L. was supported by the Bolyai János Research Scholarship of the Hungarian Academy of Sciences and by the NKFIH grant ÚNKP-23-5-BME-448.

## Declaration of conflicts of Interest

The authors declare no conflicts of interest.

## Acknowledgments

The research reported in this paper and carried out at Wigner Research Centre for Physics is supported by the infrastructure of the Hungarian Academy of Sciences.